\documentclass[preprintnumbers,prd,showpacs,floatfix,superscriptaddress,nofootinbib,twocolumn, letterpaper]{revtex4-1}
\pdfoutput=1

\usepackage{longtable}
\usepackage{bm}
\usepackage{relsize}
\usepackage{amsfonts}
\usepackage{amsmath}
\usepackage{amssymb,epsf}
\usepackage{latexsym}
\usepackage{graphicx,epsfig}
\usepackage{amssymb}
\usepackage{float}
\usepackage{subfigure}
\usepackage{epstopdf}
\usepackage[colorlinks=true,citecolor=blue,linkcolor=blue,urlcolor=black]{hyperref}
\usepackage{dcolumn}
\usepackage{psfrag}
\usepackage{wrapfig}
\usepackage{makeidx}
\usepackage{epsf}
\usepackage{color}
\usepackage{multirow}
\usepackage{mathtools}
\usepackage{tikz}

\newcommand{\xmark}{%
\tikz[scale=0.23] {
    \draw[line width=0.7,line cap=round] (0,0) to [bend left=6] (1,1);
    \draw[line width=0.7,line cap=round] (0.2,0.95) to [bend right=3] (0.8,0.05);
}}

\begin{document}

\title{Observational properties of hot-spots orbiting relativistic fluid spheres}

\author{Hanna Liis Tamm}
\email{hanna.liis.tamm@ut.ee}
\affiliation{Institute of Physics, University of Tartu, W. Ostwaldi 1, 50411 Tartu, Estonia}

\author{João Luís Rosa}
\email{joaoluis92@gmail.com}
\affiliation{Institute of Physics, University of Tartu, W. Ostwaldi 1, 50411 Tartu, Estonia}
\affiliation{University of Gda\'{n}sk, Jana Ba\.{z}y\'{n}skiego 8, 80-309 Gda\'{n}sk, Poland}

\date{\today}

\begin{abstract} 
In this work we analyze the observational properties of relativistic fluid spheres when orbited by isotropically emitting sources, known as hot spots. We consider fluid star configurations in four different regimes of compacticity, from the Buchdahl limit to non-ultra compact solutions, thus obtaining fluid stars with qualitatively different geodesic structures and observational properties. We show that the observational properties for fluid stars at the Buchdahl limit are qualitatively similar to the ones for the Schwarzschild black hole, whereas for more dilute configurations one can find observational properties similar to other fundamentally different models e.g. bosonic star configurations with and without self interactions. For solutions with a radius in the range $2.25M<R<3M$, where $M$ is the total mass of the fluid star, the presence of a light-ring (LR) pair, which becomes degenerate at $R=3M$, leads to the appearance of additional observational signatures e.g. secondary images and LR contributions, which allow one to distinguish these models from their black-hole counterparts. Fluid star configurations supported by thin shells are also analyzed and it is proven that the stability of the inner LR is increased for these solutions, resulting in a non-differentiable extremum in the effective potential of the photons. Our results suggest that compact fluid star configurations provide a suitable and physically relevant alternative to the black-hole scenario in accordance with the current generation of observational experiments.
\end{abstract}


\maketitle

\section{Introduction}\label{sec:intro}

The understanding of the gravitational interaction in its strong field regime has greatly improved in recent years due to the contribution of numerous high-precision experiments, e.g. the LIGO/Virgo gravitational wave detectors \cite{LIGOScientific:2020ibl,LIGOScientific:2016aoc,LIGOScientific:2021djp}, long baseline interferometer networks such as the Event Horizon Telescope \cite{EventHorizonTelescope:2019dse,EventHorizonTelescope:2022wkp,EventHorizonTelescope:2020qrl} and the GRAVITY instrument of the European Southern Observatory \cite{GRAVITY:2020gka,GRAVITY:2020lpa}. This rapid development of the experimental side of gravitational physics has finally provided a suitable framework to address a wide range of currently unsolved issues in physics \cite{Barack:2018yly}. In particular, one can analyze the validity of the Kerr hypothesis, which states that the end state of a complete gravitational collapse in an adequate astrophysical setting is a rotating and electrically neutral black hole spacetime \cite{Kerr:1963ud,Oppenheimer:1939ue}.

Although the observations enumerated above seem to be in close agreement with the theoretical predictions for black-hole spacetimes \cite{Will:2014kxa,Yagi:2016jml,Luminet:1979nyg,Falcke:1999pj,Gralla:2020srx}, these spacetimes are known to be prone to fundamental physical and mathematical inconveniences. An unavoidable feature of a complete gravitational collapse if the formation of a singularity \cite{Penrose:1964wq,Penrose:1969pc}, associated with a geodesic incompleteness of the spacetime. Furthermore, the cosmic censorship conjecture states that these singularities should be hidden behind event horizons, i.e., one-directional membranes directly associated with the loss of predictability and information upon the gravitational collapse \cite{Hawking:1976ra}. Due to these unwanted features, a wide variety of alternatives to the black-hole scenario, known as Exotic Compact Objects (ECOs) have been proposed \cite{Cardoso:2019rvt}, and their observational features have been widely tested \cite{Vincent:2015xta,Rosa:2022tfv,Guerrero:2022msp,Rosa:2022toh,Rosa:2023qcv}, successfully showing that these models feature observational properties similar to those of black holes, while predicting additional testable observational imprints \cite{Cardoso:2016oxy,Cardoso:2017cqb,Postnikov:2010yn,Cardoso:2017cfl,Cardoso:2016rao,Olmo:2023lil}.

Among the several types of ECO models proposed, we are interested in a family of models belonging to the class of relativistic fluid spheres \cite{Buchdahl:1959zz,Raposo:2018rjn,Cardoso:2015zqa}. In recent works, a model of a relativistic fluid sphere supported by a thin-shell (of which the Schwarzschild star is a particular case) was shown to fulfil several requirements for physical relevance, namely non-exoticity of matter and linear stability against radial perturbations \cite{Rosa:2020hex}, and existence of a shadow-like feature in accretion disk observations \cite{Rosa:2023hfm}. Furthermore, these models can have a compactness arbitrarily close to that of a black-hole, thus featuring several similar properties, e.g., the existence of an Innermost Stable Circular Orbit (ISCO) and one (or more) Light-Ring(s) (LR). Such a similarity between the relativistic fluid supported by a thin-shell and the black-hole scenario, while preserving the regularity of the spacetime, motivates further study of these configurations and their observational properties.

Several high-precision numerical codes have been developed with the purpose of simulating the observational properties of astrophysical systems. In particular, the ray-tracing software GYOTO \cite{Vincent:2011wz} was proven useful in the imaging of black-hole spacetimes in suitable astrophysical environments, e.g., the galactic centre \cite{Vincent:2020dij,Lamy:2018zvj,Vincent:2016sjq}, as well ECO models, particularly self-gravitating fundamental fields (bosonic stars) \cite{Rosa:2022tfv,Rosa:2022toh,Rosa:2023qcv}. We recur to the capabilities of this code to analyze the observational properties of the models of interest in this work.

This paper is organized as follows. In Sec. \ref{sec:theory}, we introduce the model of the static, spherically symmetric, and isotropic perfect fluid star and analyze its geodesic structure; in Sec. \ref{sec:astro} we introduce the astrometric observables of interest in this work and discuss the simulation results according to those observables; and in Sec. \ref{sec:concl} we trace our conclusions. Most of the work reported in this manuscript consists of H. L. T. BSc thesis \cite{thesis}.

\section{Theory and framework}\label{sec:theory}

\subsection{Spacetime metric and matter distribution}

In this work, we are interested in studying the observational properties of spherically symmetric and static relativistic fluid spheres composed of an incompressible fluid with a constant density. Such models have been extensively studied in other publications \cite{Rosa:2020hex,Rosa:2023hfm}, and thus here we solely provide a brief review for the self-consistency of this work. These configurations can be described by a piecewise metric consisting of two regions, an interior region populated by the relativistic fluid in the range $r<r_\Sigma$, and an exterior vacuum region in the range $r>r_\Sigma$, where $r_\Sigma$ is the radius of an hypersurface $\Sigma$ separating the two regions. The line elements that describe the interior region $ds^2_-$ and the exterior region $ds^2_+$ are given in the usual set of spherical coordinates $x^\mu=\left(t,r,\theta,\phi\right)$ by
\begin{eqnarray}\label{eq:metric_int}
    ds^2_ -&=&-\frac{1}{4}\left(3\sqrt{1-\frac{2M}{R}}-\sqrt{1-\frac{2r^2M}{R^3}}\right)^2dt^2+\\
    &+&\left(1-\frac{2r^2M}{R^3}\right)^{-1}dr^2+r^2\left(d\theta^2+\sin^2\theta d\phi^2\right),\nonumber
\end{eqnarray}
\begin{eqnarray}\label{eq:metric_ext}
ds^2_+&=&-\left(1-\frac{2M}{r}\right)dt^2+\left(1-\frac{2M}{r}\right)^{-1}dr^2+\\
&+&r^2\left(d\theta^2+\sin^2\theta d\phi^2\right),\nonumber
\end{eqnarray}
respectively, where $M$ is the total mass of the star and $R$ is the radius the star would have if the entire mass of the star is distributed volumetrically. The line element $ds^2$ for the whole spacetime can ths be written in terms of the Heaviside distribution function $\Theta$ as
\begin{equation}\label{eq:piecewise}
    ds^2 = \Theta(r_\Sigma- r)\,ds^2_- + \Theta(r-r_\Sigma)\,ds^2_+.
\end{equation}
The well-known Schwarzschild fluid star \cite{Buchdahl:1959zz} corresponds to a particular case of the model above for which $r_\Sigma=R$, but a broader family of solutions can be found with $r_\Sigma<R$. 

The matter components of the interior region are described by an isotropic relativistic perfect fluid with a constant density, i.e., one can write the stress-energy tensor $T_{\mu\nu}$ in the diagonal form
\begin{equation}\label{eq:def_tab}
    T_\mu^\nu=\text{diag}\left(-\rho,p,p,p\right),
\end{equation}
where $\rho=\frac{3M}{4\pi R^3}$ is the constant energy density of the fluid, and $p=p\left(r\right)$ is the isotropic pressure of the fluid, which is a function of the radial coordinate as
\begin{equation}\label{eq:pressure}
    p\left(r\right)=\rho\frac{\sqrt{1-\frac{2r^2M}{R^3}}-\sqrt{1-\frac{2M}{R}}}{3\sqrt{1-\frac{2M}{R}}-\sqrt{1-\frac{2r^2M}{R^3}}}.
\end{equation}
In the particular case of the Schwarzschild fluid star with $r_\Sigma=R$, the surface pressure $p\left(R\right)=0$ vanishes. It is also important to note that the central pressure $p_c=p\left(0\right)$ is a function of the total radius $R$. In particular, if the radius $R=R_b$, where $R_b=9M/4$ is known as the Buchdahl radius, the central pressure diverges. The limit $R\to R_b$ is thus known as the Buchdahl limit, and sets an upper boundary for the compacticity of regular Schwarzschild fluid stars.

The matching between the interior and the exterior regions is performed recurring to the so-called junction conditions \cite{darmois,Israel:1966rt}, a set of conditions the interior and exterior geometries must satisfy to guarantee that the resulting spacetime is itself a solution of the field equations of the theory under consideration. In General Relativity (GR), the junction conditions require the continuity of the induced metric $h_{ab}=g_{\mu\nu} e^\mu_a e^\nu_b$ and the extrinsic curvature $K_{ab}=e^\mu_a e^\nu_b \nabla_\mu n_\nu$, where $e^\mu_a$ are the projection tensors from the four-dimensional spacetime described by a set of coordinates $x^\mu$ into the three-dimensional hypersurface $\Sigma$ described by a set of coordinates $y^a$, $\nabla_\mu$ denotes the covariant derivatives, and $n_\nu$ is the normal vector to the hypersurface $\Sigma$. If the latter condition is violated, i.e., if the extrinsic curvature $K_{ab}$ is discontinuous across $\Sigma$, a thin shell of matter at $\Sigma$ is necessary to preserve the validity of the solution considered. 

\begin{figure}
    \centering
    \includegraphics[scale=0.35]{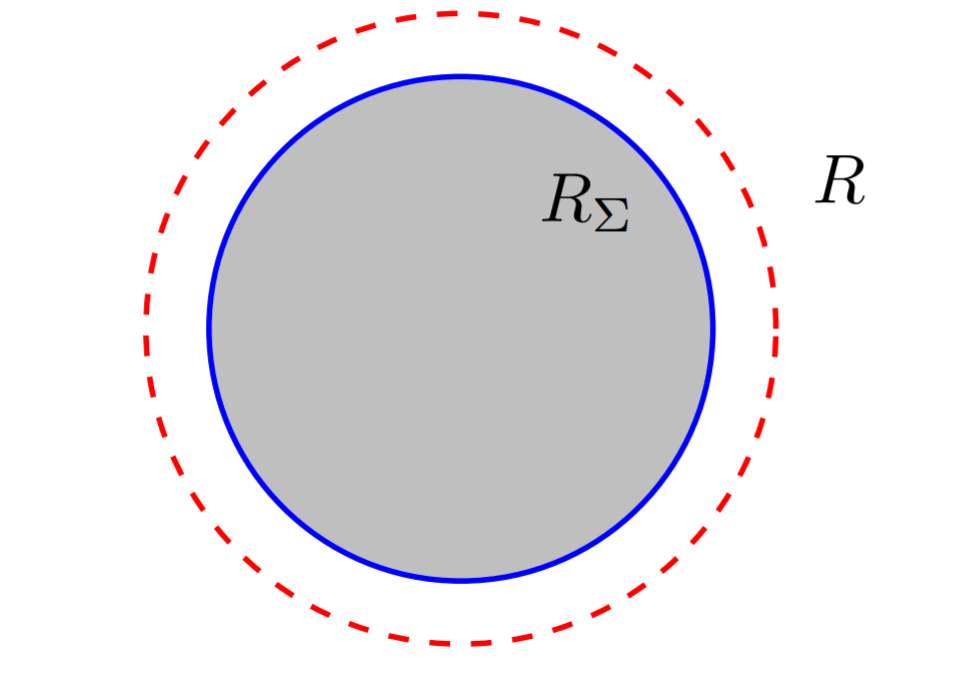}
    \caption{Schematic representation of the non-smooth relativistic fluid star model. The interior relativistic fluid region (gray) is separated from the exterior vacuum region (white) by a thin shell of matter (solid blue line) at $r=r_\Sigma<R$, where $R$ is the initial radius of the star before compression (dashed red line).}
    \label{fig:star}
\end{figure}

For the Schwarzschild star, i.e., $r_\Sigma=R$, one verifies that the metrics in Eqs.\ \eqref{eq:metric_int} and \eqref{eq:metric_ext}, as well as their Lie derivatives (and consequently their extrinsic curvatures), are continuous at $\Sigma$. Thus, the Schwarzschild fluid star features a smooth matching, i.e., without a thin shell at $\Sigma$. However, taking $r_\Sigma<R$, one obtains a class of models for which the matching is no longer smooth, i.e., a thin-shell of matter arises at $\Sigma$. These models can be interpreted as an initial Schwarzschild fluid star whose outer layers have been compressed from the initial radius $r=R$ into a smaller radius $r=r_\Sigma$, see Fig. \ref{fig:star}. In previous works \cite{Rosa:2020hex,Rosa:2023hfm}, the physical relevance of this family of models was explored. These models were shown to: (i) allow for compacticities arbitrarily close to that of a black hole without the development of an interior pressure singularity, thus not being covered by the Buchdahl limit; (ii) satisfy all of the energy conditions for a wide range of the parameter space; (iii) be stable against radial perturbations for a wide range of the parameter space; (iv) produce observable shadow-like features when surrounded by optically-thin accretion disks. These results indicate the suitability of these models as valid candidates for black-hole mimickers and thus motivates a deeper study of their observational properties.

\subsection{Geodesic motion and light rings}

The equations of motion for test particles in a background spacetime geometry can be obtained via the Lagrangian formalism, where the Lagrangian density is given by $\mathcal L=g_{\mu\nu}\dot x^\mu\dot x^\nu=-\delta$, where a dot denotes a derivative with respect to the affine parameter along the geodesics, and $\delta$ is a constant that is either $\delta=1$ for massive (timelike) test particles or $\delta=0$ for massless (null) test particles. Given that the spacetimes under study are spherically symmetric, the analysis of the Lagrangian mechanics can be simplified via the restriction to the equatorial plane $\theta=\pi/2$ and $\dot\theta=0$ without loss of generality. Under this assumption, one can define two conserved quantities, namely the energy and angular momentum per unit mass $E=-g_{tt}\dot t$, and $L=r^2\dot\phi$, respectively. Following these definitions, the radial component of the equation of motion can be written as
\begin{equation}\label{eq:eom_geodesic}
\dot r^2=\frac{V\left(r\right)}{\sqrt{-g_{tt}g_{rr}}},
\end{equation}
where $V(r)$ is the effective potential
\begin{equation}\label{eq:def_potential}
V\left(r\right)=E^2+g_{tt}\left(\frac{L^2}{r^2}+\delta\right),
\end{equation}
thus taking a similar form to the equation of motion of a particle moving along a one-dimensional potential function $V\left(r\right)$. 

Circular orbits are defined by $\dot r=\ddot r=0$ which, from Eq.\ \eqref{eq:eom_geodesic}, implies that $V\left(r\right)=V'\left(r\right)=0$. In particular, circular null geodesics, commonly known as Light Rings (LRs), can be found by further taking $\delta=0$. In this setup, it is practical to perform a redefinition of the potential function in the form
\begin{equation}\label{eq:redef_potential}
\bar V\left(r\right)=\frac{V\left(r\right)}{L^2}-\frac{1}{b^2},
\end{equation}
where we have defined $b=L/E$ as the impact parameter. The stationary points of $\bar V$ correspond to the radii of the LRs of the background spacetime, if any. 
The stability of the orbits can be analyzed via the second-order derivative of $V$, for which the orbits are stable whenever $\bar V''(r) > 0$ and unstable otherwise.

\begin{figure}
    \centering
    \includegraphics[width=0.49\textwidth]{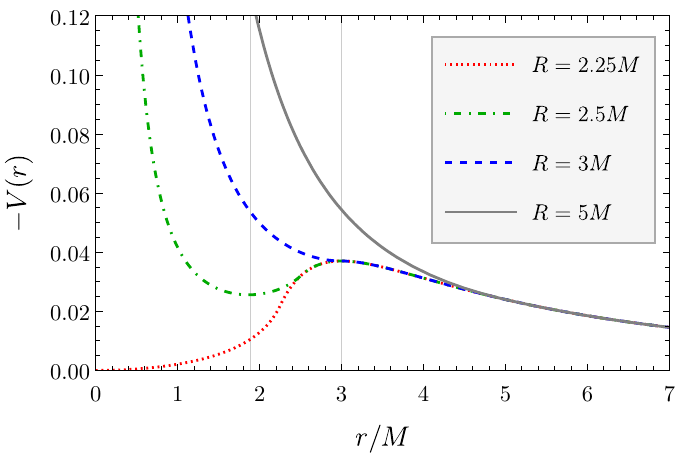}
    \caption{Effective potential $-\bar V(r)$ defined in Eq. (\ref{eq:redef_potential}) for $r_\Sigma =R$, $R\in\{2.25M, 2.5M, 3M,5M\}$. The vertical lines correspond to extrema of the potential at $\bar r_{LR}=1.89M, r_{LR}=3M$.}
    \label{fig:eff_pot_1}
\end{figure}

\begin{figure}
    \centering
    \includegraphics[width=0.49\textwidth]{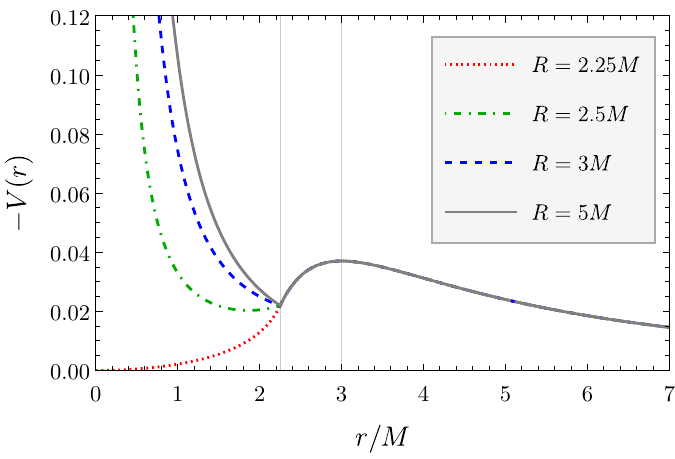}
    \caption{Effective potential $-\bar V(r)$ defined in Eq. (\ref{eq:redef_potential}) for $r_\Sigma =2.25M$, $R\in\{2.25M, 2.5M, 3M,5M\}$. The vertical lines correspond to extremum of the potential at $r_{LR}=3M$ and to the radius of the shell $r_\Sigma=2.25M$, which in this case also corresponds to the LR $\bar r_{LR}=2.25M$ for the configurations with $R=\{3M,2.5M\}$.}
    \label{fig:eff_pot_2}
\end{figure}

According to a recent theorem published in Ref. \cite{Cunha:2017qtt}, in regular ultra-compact spacetimes LRs come in (possibly degenerate) pairs of a stable and an unstable LR. Indeed, the fluid spheres investigated in this work can exhibit this phenomena for certain ranges of the star radius $R$. Choosing a radius greater than the Buchdahl limit $R \geq 9M/4$, the following behaviour is exhibited:
\begin{enumerate}
    \item If $R = 9M/4$, the condition $\bar V'(r) = 0$ is satisfied for two values of $r$: (a) $r_{LR} = 3M$, representing an unstable LR, and (b) $r=0M$, representing a stable potential well.
    \item In the parameter range $9M/4 < R \leq 3M$, the condition $V'(r) = 0$ produces two solutions corresponding to a pair LRs - one stable, the other unstable. The unstable LR with $r_{LR} = 3M$ originates from the well studied exterior vacuum solution, as expected, but the interior region produces a stable LR at $r=\bar r_{LR}$. The radius of the stable LR can be found from the relation
    \begin{equation}\label{eq:stable_LR_radius}
        \bar r_{LR} = R\sqrt{\frac{R}{R_b}}\sqrt{\frac{R-R_b}{R-2M}}.
    \end{equation}
    Note that $0 < \bar r_{LR} \leq 3M$ for star radius values $R \geq 9M/4$ and in the case $R =3M$, the stable LR coincides with the unstable LR at $r_{LR} = \bar r_{LR} = 3M$, thus producing a degenerate pair.
    \item If $R > 3M$, no solutions are found to satisfy $\bar V'(r) = 0$, and therefore no LRs are present in the spacetime.
\end{enumerate}
The effective potential $\bar V(r)$ is plotted in Fig. \ref{fig:eff_pot_1} for the case of the Schwarzschild fluid star, i.e. $r_\Sigma = R$.
The minimum in the potential function, which appears in the range $9M/4 < R < 3M$, merges with the maximum at $R = 3M$, producing a saddle point. However, for cases where $r_\Sigma < R$, the effective potential features additional noteworthy points at the shell radius $r_\Sigma$,
increasing the stability of the interior stable LR or producing additional stable LRs at the shell radius $r_\Sigma$ (see Fig. \ref{fig:eff_pot_2}).
\color{black}

\section{Astrometry and observables}\label{sec:astro}

Let us now consider the observational properties of fluid stars orbited by hot spots. To analyze this astrophysical setup, we recur to the open-source software GYOTO \cite{Vincent:2011wz}, where we model the hot spot as an isotropically emitting spherical source, moving along a circular orbit of a given orbital radius $r_o = \{8M, 10M, 12M\}$ restricted to the equatorial plane $\theta=\pi/2$. We take the radius of the hot spot to be $r_H=M/2$, where M is the total mass of the fluid star, and we set the observer's distance from the center of the system at $r=1000M$. The observer lies above the equatorial plane with some inclination angle $\theta=\{20^\circ,50^\circ,80^\circ\}$. The software performs a ray-tracing of the setup described and outputs a two-dimensional matrix of specific intensities $I^\nu_{lm}$ for different time instants $t_k$ in the range $t_k\in\left[0,T\right[$, where $T$ is the orbital period of the source. The resulting cube of data $I_{klm}=\Delta\nu I^\nu_{lm}$, where $\Delta\nu$ is the spectral width, can then be sued to produce several observable quantities, e.g., the time integrated fluxes $\left<I\right>_{lm}$, the temporal fluxes $F_k$, and the temporal centroids $\vec{c}_k$, which are defined as
\begin{equation}\label{eq:intflux}
\left<I\right>_{lm}=\sum_k I_{klm},
\end{equation}
\begin{equation}\label{eq:timeflux}
F_k=\sum_{l,m}\Delta\Omega I_{klm},
\end{equation}
\begin{equation}\label{eq:timecent}
\vec{c}_k=\sum_{l,m}\Delta\Omega I_{klm} \vec{r}_{lm},
\end{equation}
where $\Delta\Omega$ is the solid angle of a single pixel and the vector $\vec{r}_{lm}$ represents the displacement of the pixel $\{l,m\}$ with respect to the central point of the observed image. Finally, one can construct the temporal magnitude $m_k$ from the temporal fluxes $F_k$ as
\begin{equation}\label{eq:magnitude}
m_k=-2.5 \log\left(\frac{F_k}{\min F_k}\right).
\end{equation}

To analyze the behaviour implied by the effective potential, we consider four Schwarzschild fluid star configurations described in Table \ref{tab:simulated_models}, including a comparison with a Schwarzschild BH. The BH and models M1-M3 are ultra-compact objects (i.e., they feature at least one LR in the spacetime), of which the Schwarzschild BH and model M1 feature singularities in the metric component and the central pressure, respectively. The model M4 represents a fluid star with a smaller compacticity, which does not feature neither LRs nor singularities.
\begin{table}
    \centering
    \begin{tabular}{c|c c c c c}
         & $R /M$ & $r_\Sigma /M$ & LR & singularity & EH \\ \hline 
        Schwarzschild BH & 0 & 0 & \checkmark & \checkmark & \checkmark\\ 
        M1 & 2.25 & 2.25 & \checkmark & \checkmark & \xmark\\ 
        M2 & 2.5 & 2.5 & \checkmark & \xmark& \xmark\\ 
        M3 & 3 & 3 & \checkmark & \xmark& \xmark\\ 
        M4 & 5 & 5 & \xmark & \xmark & \xmark 
    \end{tabular}
    \caption{Spacetime properties of the models considered and in comparison with the Schwarzschild BH, namely the normalized star radius $R/M$, the normalized shell radius $r_\Sigma/M$ and the presence of LRs, singularities, and EHs. We note that the singularities in the BH and M1 models are of a different nature: the first being a curvature singularity and the second being a pressure singularity.}
    \label{tab:simulated_models}
\end{table}
The hot spot is imaged at 180 orbital positions for a resolution of $500\times 500$ pixels, and the time integrated image of the flux, the temporal magnitudes, and temporal centroids are produced. We note that no configurations with $r_\Sigma\neq R$ were considered due to software limitations, as GYOTO is not capable of ray-tracing in background spacetimes featuring discontinuous Christoffel symbols. For an analysis of the observational effects of the thin-shell using a different software, we refer to Ref. \cite{Rosa:2023hfm}.

\subsection{Integrated flux}
\begin{figure*}
    \centering
    \includegraphics[width=\textwidth]{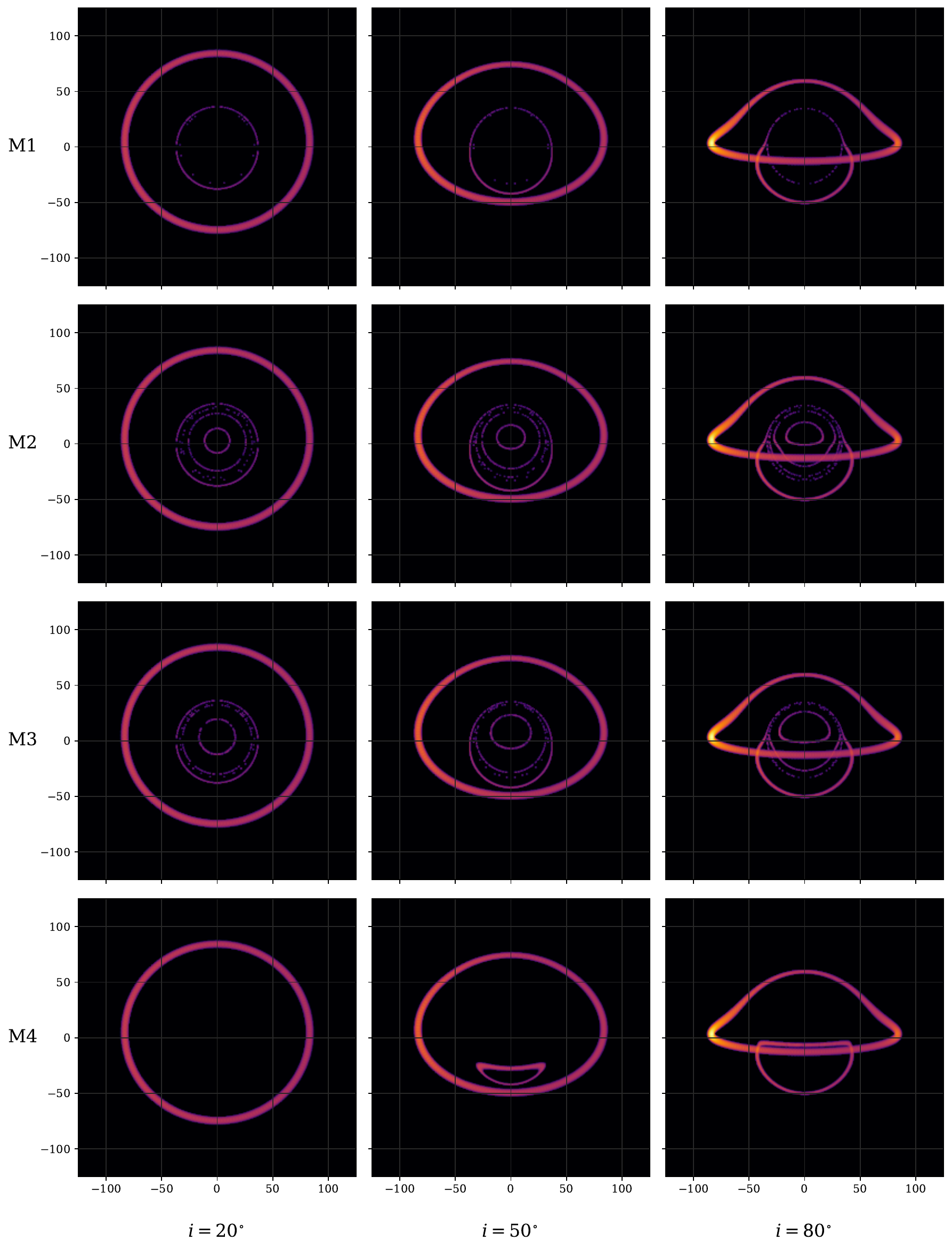}
    \caption{Time integrated images of models M1-M4 for inclination angles $i \in \{20^{\circ},50^{\circ},80^{\circ}\}$ at a hot spot orbital radius of $r_o = 12M$. Models M1 and M4 resemble a Schwarzschild BH and a boson star, respectively, while models M2-M3, corresponding to ultra-compact horizonless objects, feature novel trajectories in the interior. The $x$- and $y$-axis are in units $\mu a s$.}
    \label{fig:all_int}
\end{figure*}

The integrated flux for the models M1-M4 is shown in Fig.\ \ref{fig:all_int} at a hot spot orbital radius of $r_o=12M$ and inclination angles of $i=\left\{20^\circ,50^\circ,80^\circ\right\}$ (for radii $r_o = 8M$, $r_o = 10M$, see Appendix \ref{sec:appendixA}).

The first row in Fig.\ \ref{fig:all_int} depicts the integrated flux for the model M1, representing the Buchdahl limit $r_\Sigma = R = 2.25M$, for three inclinations. The integrated flux consists of three tracks: a) a primary track (exterior track), representing photons which are deflected less than $90^{\circ}$ around the central object and do not cross the equatorial plane after emission; b) a secondary (in-between) track, corresponding to photons which are deflected at least $90^{\circ}$, but not enough to form loops, thus crossing the equatorial plane once after emission; c) a LR track (interior), representing photons that perform closed loops, crossing the equatorial plane at least twice after emission.
As the inclination angle increases, the primary track distorts along the $x$-axis and part of the secondary track moves away from the LR track, moving towards the negative of the $y$-axis. As the hot-spot radius $r_o$ increases, the width of the primary and secondary track increases, while the LR track remains virtually unchanged (see Fig.\ \ref{fig:2p25} in Appendix \ref{sec:appendixA}). 
These properties indicate that the model is qualitatively identical to the classic BH based on the exterior Schwarzschild solution, while small quantitative differences can be observed in the magnitude of the light deflection, comparing with \cite{Rosa:2022toh}.

The second model M2 shows novel features, as seen in the second row of Fig.\ \ref{fig:all_int}. In addition to the primary, secondary and the LR tracks, the integrated images show multiple tracks in the interior, corresponding to additional LR induced tracks, with the exception of the innermost track, which is a secondary track. Increasing the inclination angle, the lensed primary and secondary tracks are deformed similarly to the previous case. However, the lensing of the interior tracks can also be observed: the innermost secondary track is affected the most, while the LR tracks are only slightly affected. Indeed, the only LR track that remains virtually unchanged is the one corresponding to the outer unstable LR, that was present also for the model M1. As the orbital radius of the hot spot increases, the lensed primary and secondary tracks expand similarly to BHs and the additional trajectories in the interior decrease in size (see Fig.\ \ref{fig:2p5}).
The time-integrated images of the model M3 in the third row of Fig.\ \ref{fig:all_int} show the same components as the model M2, including the primary and secondary track as well as interior tracks: LR induced tracks and a secondary track which appears closest to the center. However, this model features one less LR track compared to the previous model M2, producing a qualitative difference. In addition, the LR tracks and the innermost secondary track have an increased diameter compared to the previous case. Note that while this case corresponds to a degenerate pair of LRs, only the outer LR track remains unchanged for variations of the observation angle.

The model M4 in the final row of Fig.\ \ref{fig:all_int} consists of the primary, secondary and the "plunge-through" track, the latter defined as the photons reaching the observer by crossing the interior of the compact object through a region that would correspond to the EH in the BH counterpart. This model thus resembles a compact object without a LR, e.g., a boson star without self-interactions, which has been studied in previous works. In this case, we observe an angle-dependent qualitative behaviour: the presence of the secondary and plunge-through tracks depends on the inclination angles and orbital radii (see Fig.\ \ref{fig:5}), while for the previous models M1-M3 the number of tracks remains constant throughout changes in the inclination angle and orbital radius.

\begin{figure}
    \centering
    \includegraphics[width=0.49\textwidth]{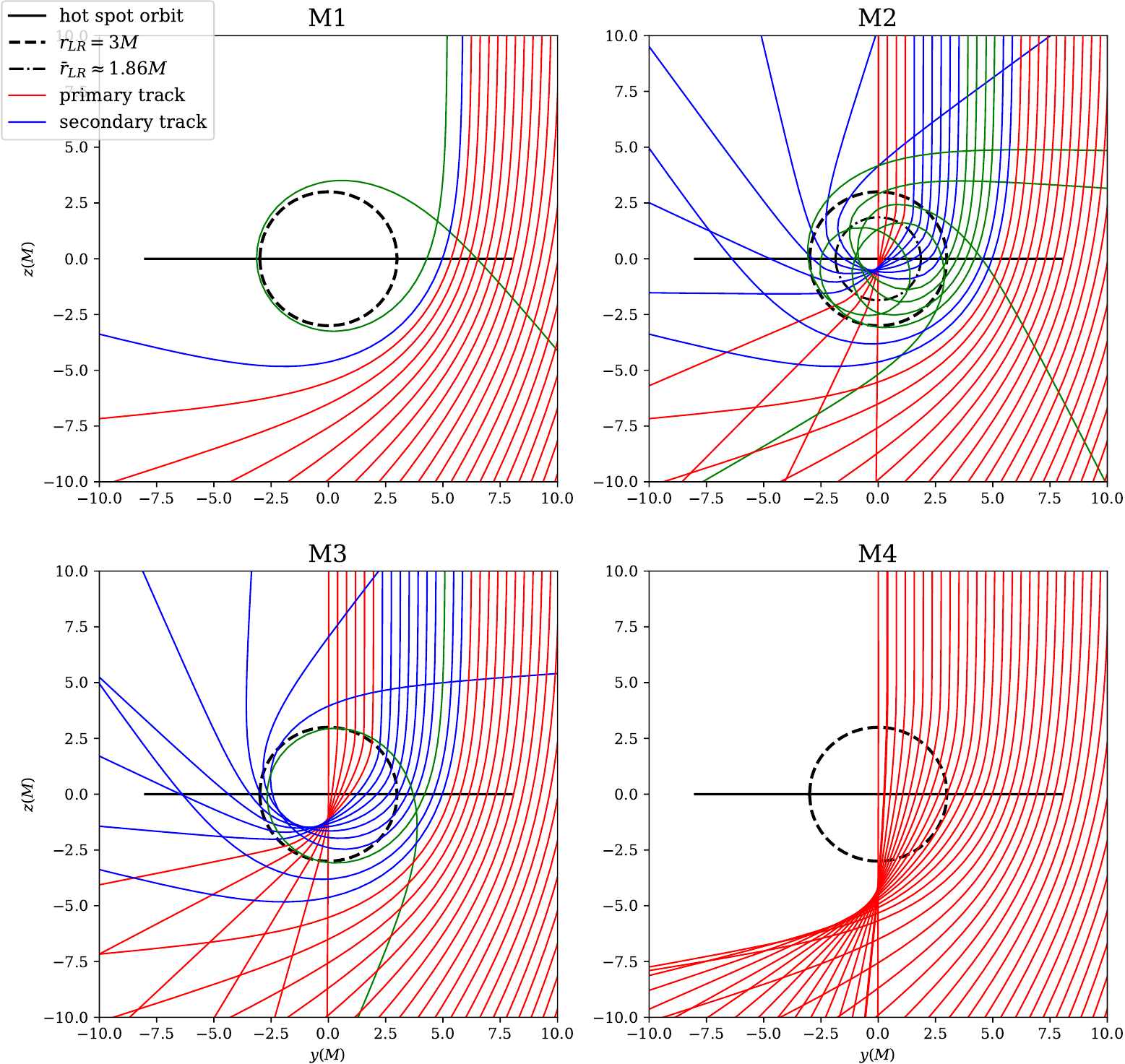}
    \caption{The geodesics of photons for models M1-M4 for $i = 0.01^{\circ}$. The dashed line represents the unstable LR with $r_{LR} = 3M$, the dot-dashed line represents the stable LR with $\bar r_{LR}$ and the solid black line represents the position of the equatorial plane for $i=0.01^{\circ}$. The observer is located towards the top of the image at $z = 1000M$; the photons are traced on the line $y = 1000M$ and their trajectories correspond to primary (red), secondary (blue) or LR photons (green).}
    \label{fig:geodesics}
\end{figure}

For each configuration, consider the geodesic congruences shown in Fig.\ \ref{fig:geodesics}, depicting photons that are deflected less than $90^\circ$ in red (those correspond potentially to a primary track), photons deflected more than $90^\circ$ but not enough to produce loops in blue (those correspond potentially to secondary tracks), and LR photons in green. 
The solid black line in Fig.\ \ref{fig:geodesics} represents the orbital plane of the hot spot for $r_o = 8M$ and $i = 0.01^{\circ}$ (an inclination angle of exactly $i=0^\circ$ is not supported by the software). The observational properties, i.e., the number and type of the photon tracks in the observed image, can be inferred for any inclination angle and hot spot orbital radius by varying the angle and length of the solid black line.

For the model M1, photons with an impact parameter $b$ larger than the critical impact parameter $b_c = 3\sqrt{3}M$, are deflected slightly and form either the primary or secondary track (the red and blue trajectories). Photons closer to the center with an impact parameter close to $b_c$ perform loops around the center and therefore represent LR tracks (the green trajectory). Any photon with an impact parameter below $b_c$ is captured due to the potential well in the center of the star (see Fig.\ \ref{fig:eff_pot_1}), which explains the absence of observable photons in the interior.
For the following models M2-M4, the absence of the potential well permits additional photon trajectories, which cross inside the radius of the LRs. Regarding the first three models M1-M3, the geodesic behaviour of photons with $b \geq b_c$ follows the exterior of a classic BH, exhibiting primary, secondary and LR tracks on the exterior. However, for the models M2-M3, the interior structure changes: photons with $b < b_c$ are allowed to escape and form either LR, secondary or primary tracks, depending on the impact parameter. In particular, as the impact parameter of a photon decreases, its gravitational deflection decreases until it passes through the center of the star with no deflection. The degeneracy of the LR is well illustrated in the comparison between the second and third model - in model M2, photons on LR tracks perform loops influenced by both the exterior unstable LR with $r_{LR} = 3M$ and interior stable LR with $\bar r_{LR}$, whereas for the model M3, the trajectories follow the course of only the degenerate LR radius $r_{LR} = \bar r_{LR} = 3M$. Note that at certain points on the horizontal line (corresponding to a certain hot-spot orbital radius), blue lines are able to intersect. These points represent configurations in which the hot spot is lensed into multiple secondary tracks, which are visible simultaneously. Due to the nature of the secondary tracks, it becomes evident that for any orbital radius $r_o$, two secondary tracks are visible simultaneously, one originating from the exterior and the other from the interior, the latter thus corresponding to a sort of plunge-through image.
The model M4 represents a non-ultra compact case where no secondary nor LR tracks appear due to the small compacticity of the star. For the inclination $i=0.01^{\circ}$, every geodesic crosses the equatorial plane only once, and thus only the primary image is present for this inclination.

\begin{figure}
    \centering
    \includegraphics[width=0.49\textwidth]{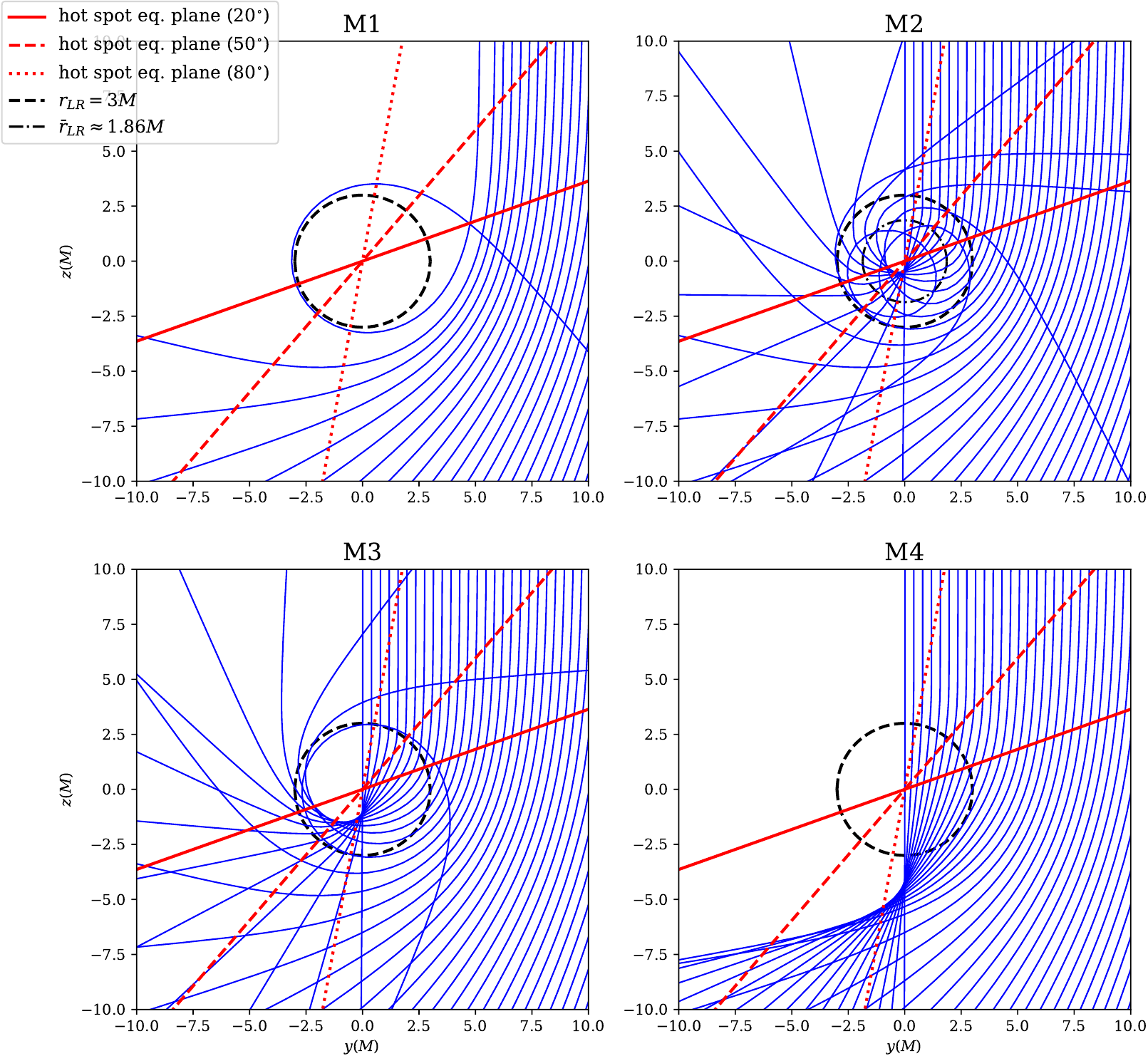}
    \caption{The geodesics of photons for models M1-M4. The dashed black line represents the unstable LR with $r_{LR} = 3M$, while the dash-dotted black line represents the stable LR with $\bar r_{LR}$. The solid, dashed and dotted red lines represent the position of the equatorial plane of the hot spot for $i\in \{20^{\circ}, 50^{\circ},80^{\circ}\}$, respectively. The observer is located towards the top of the image at $z = 1000M$; the photons are traced on the line $y = 1000M$.}
    \label{fig:geodesics_all}
\end{figure}

Let us consider the congruence for the other observation inclinations considered in this work, i.e., $i=\left\{20^\circ,50^\circ,80^\circ\right\}$, which by symmetry are equivalent to inclining the equatorial plane in the figures of the geodesic congruences, represented in red in Fig.\ \ref{fig:geodesics_all}. Note that the type of each trajectory is not uniquely defined. As the inclination of the equatorial plane changes, it may happen that the number of times a given trajectory crosses the equatorial plane varies. The first model M1 matches the Schwarzschild case: all geodesics are projected into the observer's screen regardless of the inclination. This implies that the primary, secondary, and LR tracks are always present, independently of the observer's inclination. The latter also holds true for models M2-M3. However, the congruence of model M4 exhibits a behaviour dependent on the inclination and orbital distance: for smaller inclinations, only trajectories corresponding to a primary track intersect the equatorial plane. Starting from a certain critical inclination and critical hot-spot orbital radius, one can observe that some geodesics cross the equatorial plane once after emission and before reaching the observer. For these inclinations, an additional secondary track appears, corresponding to the secondary and plunge-through images. This behavior is evident in the bottom row of Fig.\ \ref{fig:5}.

\subsection{Magnitudes}
\begin{figure*}
    \centering
    \includegraphics[width=\textwidth]{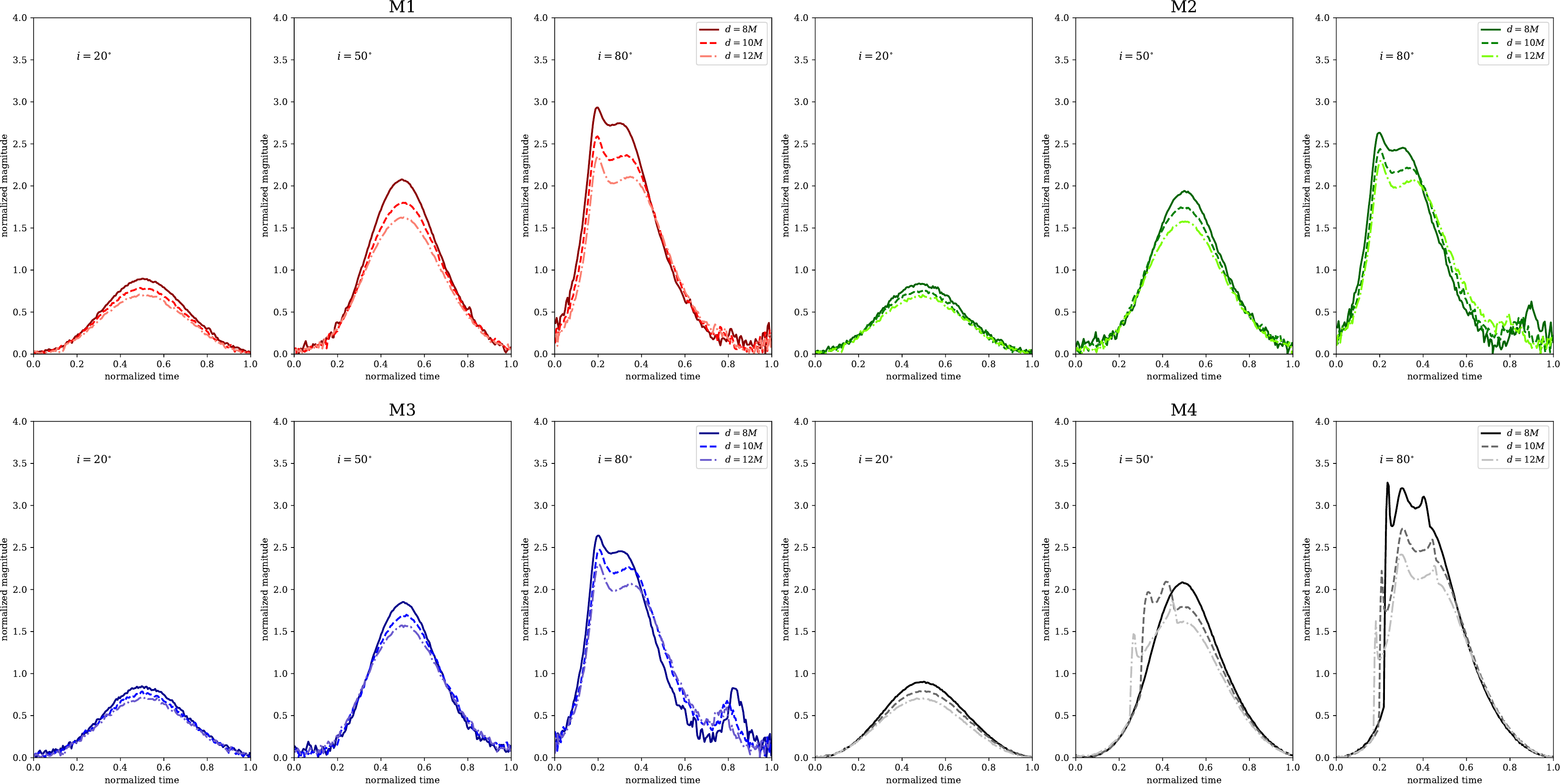}
    \caption{Temporal magnitudes $m_k$ of the hot spot for models M1-M4 for inclination angles $i\in \{20^{\circ},50^{\circ},80^{\circ}\}$ and orbital radii $r_o \in \{8M,10M,12M\}$. See text for details.}
    \label{fig:magnitudes}
\end{figure*}

The temporal magnitudes $m_k$ for a full orbit, as defined in Eq.\ \ref{eq:magnitude}, are shown in Fig.\ \ref{fig:magnitudes}.
Consider the model M1 first, shown in the top left of Fig.\ \ref{fig:magnitudes} in red. For inclinations $i=20^{\circ}$ and $i = 50^{\circ}$, the temporal magnitude takes the form of a central peak due to the relativistic Doppler effect, whereas for the case $i=80^{\circ}$, two additional peaks can be observed: one peak, which is higher in magnitude and appears slightly before the central peak; and one lower peak near the end of the orbital period. These peaks are caused by the beaming of the secondary track, which increases with inclination. Indeed, the luminosity of the secondary image is approximately constant for low inclinations, but it suffers a strong beaming for high inclinations when the source moves behind the central object (see Fig.\ \ref{fig:2p25} in Appendix \ref{sec:appendixA}). Decreasing the orbital distance or increasing the inclination increases the magnitude of the hot spot  due to its increased orbital velocity and, consequently, stronger relativistic Doppler and beaming effects are produced. One can also observe slight numerical noise, which is caused by the presence of LRs. Due to the limited resolution of the image, the LR photons are not always present in the observed imaged throughout the whole orbit, which causes slight increases and decreases in luminosity. Since the magnitude is a logarithmic quantity, the contribution from the LR becomes larger at smaller magnitudes, increasing the noise in the low flux region.

The models M2 and M3 depicted in Fig.\ \ref{fig:magnitudes} in green and blue, respectively, feature two types
of contributions, one corresponding to additional LR tracks which form in the interior, and the other to the additional secondary image closest to the center of the star. In the temporal sequence, the LR tracks appear one after the other, the outermost tracks being shown at the beginning of the orbital period and the tracks closer to the center being shown as time progresses. However, since the intensity of the flux of photons on LR tracks is comparatively small, these contributions are represented
by very small deviations from the main curve of the magnitude, which cannot be distinguished from numerical noise. For the model M2, the additional secondary track in the center appears at
all times, which does not alter the curve for inclinations $i=20^{\circ}$, $i=50^{\circ}$. At higher inclinations, the strong beaming of the secondary image causes the appearance of an additional peak, similar to what was previously mentioned in model M1. For the model M3 at 80 degrees, this increase in the second peak is greater due to the increased size of the secondary images in this model in comparison with the M2 model. Therefore, comparing Schwarzschild BHs with ultra-compact horizonless objects, a new visual signature can be observed in the magnitude. In particular, the appearance of additional peaks in the temporal magnitude corresponding to new secondary type tracks in the observed image.

For the final model M4, depicted in black in Fig. \ref{fig:magnitudes}, a double hump arises, starting from a certain inclination and orbital radius. Supported by the integrated flux in Fig.\ \ref{fig:5}, these extra peaks are caused by the appearance of the secondary track, which appears as a single image (causing the first additional peak), splits into two components, one of which previously defined as a "plunge-through" image, and finally merges again into a single image before disappearing (causing the second additional peak). As the orbital distance increases, the merged track appears for a longer time period. In the case $i=80^{\circ}$, the strong beaming of the secondary image, similarly to what was mentioned for the previous models, causes the appearance of an additional peak in between the appearance and disappearance of the secondary image. Note that the magnitude for the final model is free of noise due to the absence of LRs.

\subsection{Centroid}
\begin{figure*}
    \centering
    \includegraphics[width=\textwidth]{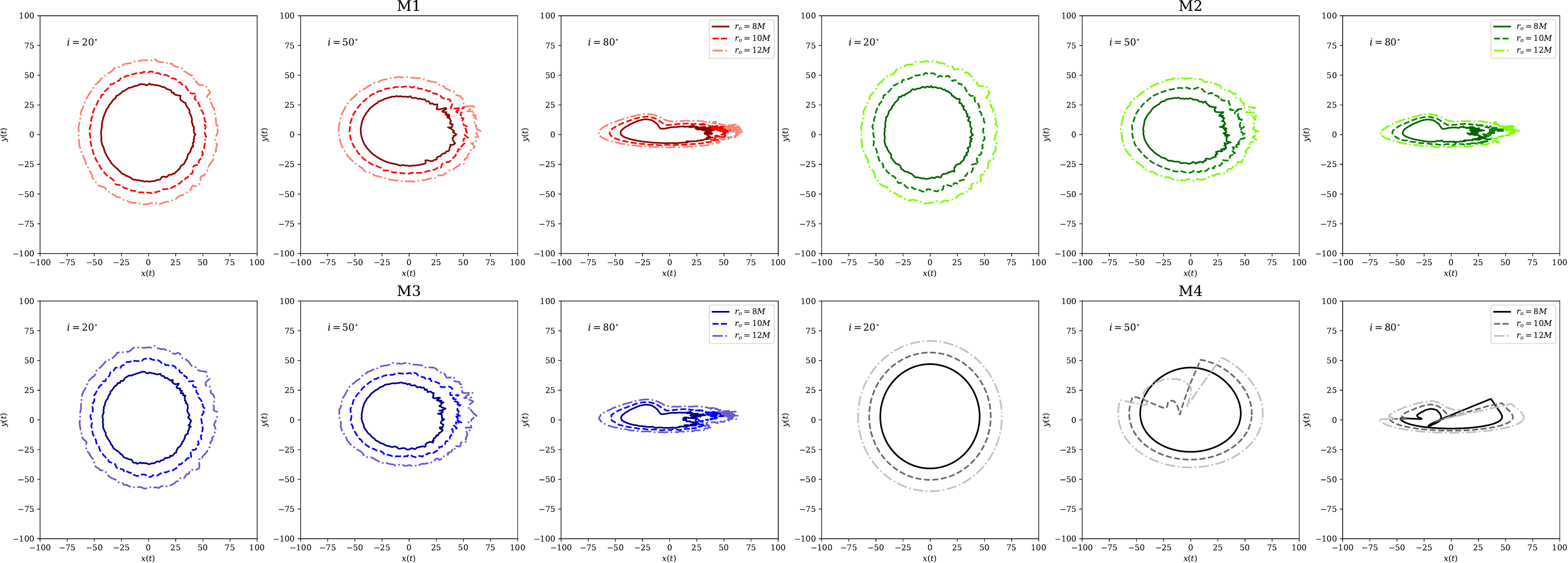}
    \caption{Centroid trajectories $\vec c_k$ during the orbital period for models M1-M4 for angles $i\in \{20^{\circ},50^{\circ},80^{\circ}\}$ and orbital radii $r_o \in \{8M,10M,12M\}$. The $x$- and $y$-axis are in units $\mu a s$. See text for details.}
    \label{fig:centroids}
\end{figure*}

The temporal centroids are shown in Fig.\ \ref{fig:centroids}, depicting the movement of the centroid  as defined in Eq.\ (\ref{eq:timecent}) during the orbital period. Note that the $x$- and $y$-axis are inverted in regard to the integrated images. As expected, increasing the orbital radius increases the size of the centroid trajectory. The centroid trajectory is approximately a circle at smaller angles and includes noise whenever LRs are present in the spacetime, for the reasons outlined in the previous section. As the inclination angle increases, the primary track becomes asymmetric and the centroid trajectory is flattened along the $y$-axis, approaching the form of an ellipse. Starting from a certain critical inclination $i_c$, the stronger beaming of the secondary image pushes the centroid to the center, thus creating a cusp in the right side of the centroid trajectory. As the primary image dominates on the left side of the trajectory, the left side resembles a distorted ellipse, whereas on the right side, the appearance of LR and secondary tracks push the centroid inwards.

Focusing on the models M2-M3, the centroid tracks feature new observables in the form of additional peaks on the right side of the centroid trajectory, caused by the additional secondary and LR tracks, which again are particularly visible for higher inclinations due to the strong beaming of the additional images. In these models, the contribution of LRs becomes more apparent, as LR tracks are situated closer to the center than the primary and secondary track, the centroid is pushed further towards the center than in the model M1. Furthermore, since LR tracks appear at different orbital positions, they are represented in the centroid track as distinct peaks. Comparing models M2 and M3, we notice a slight difference regarding the right side: for the model M3, the centroid is pushed further towards the center due to the increased dominance of the additional secondary track closest to the center.

Regarding the model M4, the noise from the LRs is absent, as expected. Furthermore, one again observes that the appearance of a secondary track depends on the observation inclination and the orbital radius, as only for certain values of these quantities above a certain critical value is the regular form of the ellipse broken with sharp dips towards the center, corresponding to the appearance and disappearance of the secondary and plunge-through images. In between those dips, the push is alleviated by the separation of the two tracks as the secondary image travels along a course further from the center than the plunge-through image. This behaviour is represented by a bump in the centroid track; however, it is not symmetrical with respect to the dips towards the center due to the Doppler shift: the intensity decreases on the right side of the integrated image. Furthermore, as the orbital distance increases, the secondary image travels further and increases the size of the bump.

\section{Conclusions}\label{sec:concl}

In this work, we studied the visual signatures of hot spots orbiting relativistic perfect fluid spheres, a class of BH mimickers, which are described by a piecewise metric composed of an interior and exterior Schwarzschild solutions, separated by a hypersurface where a thin shell might arise depending on the matching radius. As we have found, the observables of a hot spot are dependent on the presence and number of LRs in the spacetime, which in turn depend on the parameters of the configuration, namely the star radius $R$ and shell radius $r_\Sigma$. We have performed orbital simulations in the background of several distinct fluid star configurations using the ray-tracing software GYOTO, including a qualitative comparison with the Schwarzschild BH.

We investigated four different configurations, the Schwarzschild equivalent model M1 satisfying $r_\Sigma = R = 2.25M$, two ultra-compact horizonless objects M2 and M3 ($2.25M < R \leq 3M$) and a non-ultra-compact horizonless object M4 ($R > 3M$). In the densest configuration, the spacetime features an unstable LR as well as a potential well, the latter trapping photons on the inside of the object and resulting in observational properties, namely the integrated fluxes, and temporal magnitudes and centroids of the observation similar to those of a Schwarzschild BH, with the advantage of having no event horizons. As the compacticity of the configuration decreases, the spacetime features an additional stable LR, which gives rise to novel observables: additional interior tracks in the lensed images, a contribution of an additional secondary track in the magnitude and additional shifts towards the center in the centroid trajectories, caused by additional tracks in the integrated flux. For ultra-compact configurations, a specific case emerges where the unstable and stable LR coincide, producing slight differences in the integrated image (fewer LR tracks), magnitudes (increased contribution of the innermost secondary track) and centroid trajectories (centroids shifted further towards the center). Past a certain star radius, the spacetime is not compact enough to produce LRs and therefore represents a non-ultra-compact horizonless object, including a specific combination of a secondary and plunge-through track in the lensed images, which create additional peaks in the magnitude and distinct shifts towards the center in the centroid tracks.

The results obtained in this work seem to be in a close agreement with the ones obtained in previous works where bosonic star configurations with \cite{Rosa:2023qcv} and without \cite{Rosa:2022toh} self-interactions were analyzed. Indeed, ultracompact boson stars produce observational properties similar to those of the models M2 and M3, while non ultracompact boson stars feature observational properties similar to those of the model M4. These results indicate that, as long as the interaction of photons with the matter contents of the compact object are neglected, compacticity plays the most dominant role in defining the observational properties of black hole mimickers, independently of the matter fields considered as a source (perfect fluids in the cases analyzed in this work or fundamental fields in the case of the bosonic star models mentioned above). A possible exception to the conjecture stated in this paragraph is the gravastar model, as the interior time component of the metric that describes this model is fundamentally different from the ones describing perfect fluid and bosonic stars. An analysis of gravastar models is currently ongoing.

This work intends to broaden the perspective on the study of BHs and other ECOs by analyzing the motion of hot spots around compact objects. By comparing the results of ECO models with the classic BH model, the evidence for BHs can be investigated with greater precision and better-defined limits. Due to qualitative similarities between the Schwarzschild BH and the fluid sphere satisfying the Buchdahl limit, the fluid star could serve as a valid BH mimicker. However, we have also found configurations in which the observables of the hot spot feature slight differences, which could be used to distinguish fluid stars from other compact objects. Due to upcoming advances in very-long-baseline interferometry, for
example, the next generation of the Event Horizon Telescope, this distinction could potentially be observed.

\color{black}

\begin{acknowledgments}
H. L. T. and J.L.R. acknowledge the European Regional Development Fund and the programme Mobilitas Pluss for
financial support through Project No.~MOBJD647. J. L. R. further acknowledges project No.~2021/43/P/ST2/02141 co-funded by the Polish National Science Centre and the European Union Framework Programme for Research and Innovation Horizon 2020 under the Marie Sklodowska-Curie grant agreement No. 94533. We also acknowledge Fundação para a Ciência e Tecnologia through project number PTDC/FIS-AST/7002/2020.
\end{acknowledgments}

\pagebreak

\appendix

\section{Integrated images for full parameter space}\label{sec:appendixA}
For the purpose of completeness, we present the time-integrated images of the flux of the hot spot for the full parameter space $r_o = \{8M,10M,12M\}$, $i = \{20^{\circ},50^{\circ},80^{\circ}\}$.

\begin{figure*}
    \centering
    \includegraphics[width=\textwidth]{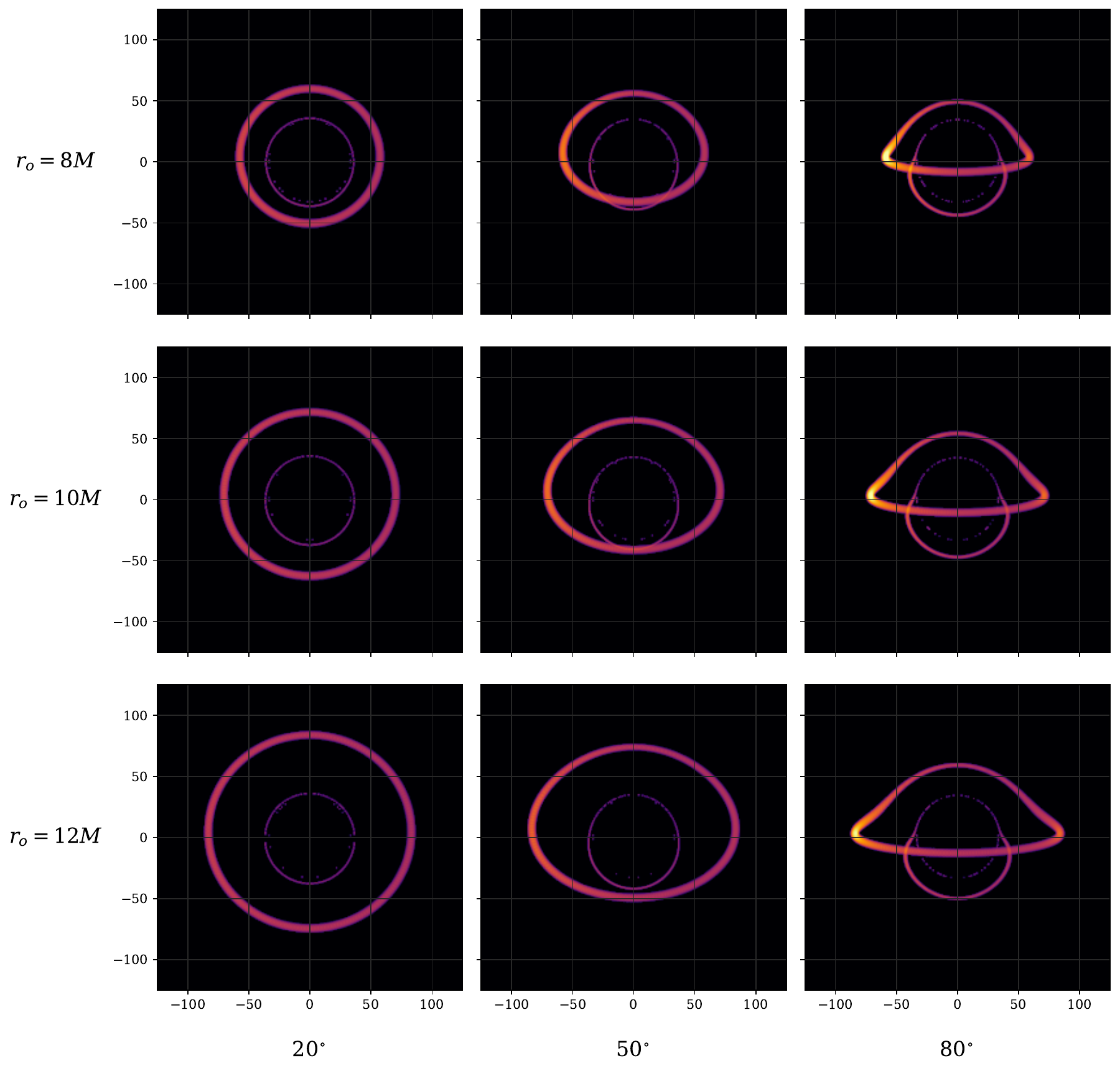}
    \caption{The integrated flux of the model M1 - $r_\Sigma = R = 2.25M$ for orbital radii $r_o=\{8M,10M,12M\}$ (increasing from top to bottom) and inclination angles $i=\{20^{\circ},50^{\circ},80^{\circ}\}$ (increasing from left to right). The $x$- and $y$-axis are in units $\mu a s$. The model resembles a Schwarzschild BH.}
    \label{fig:2p25}
\end{figure*}

\begin{figure*}
    \centering
    \includegraphics[width=\textwidth]{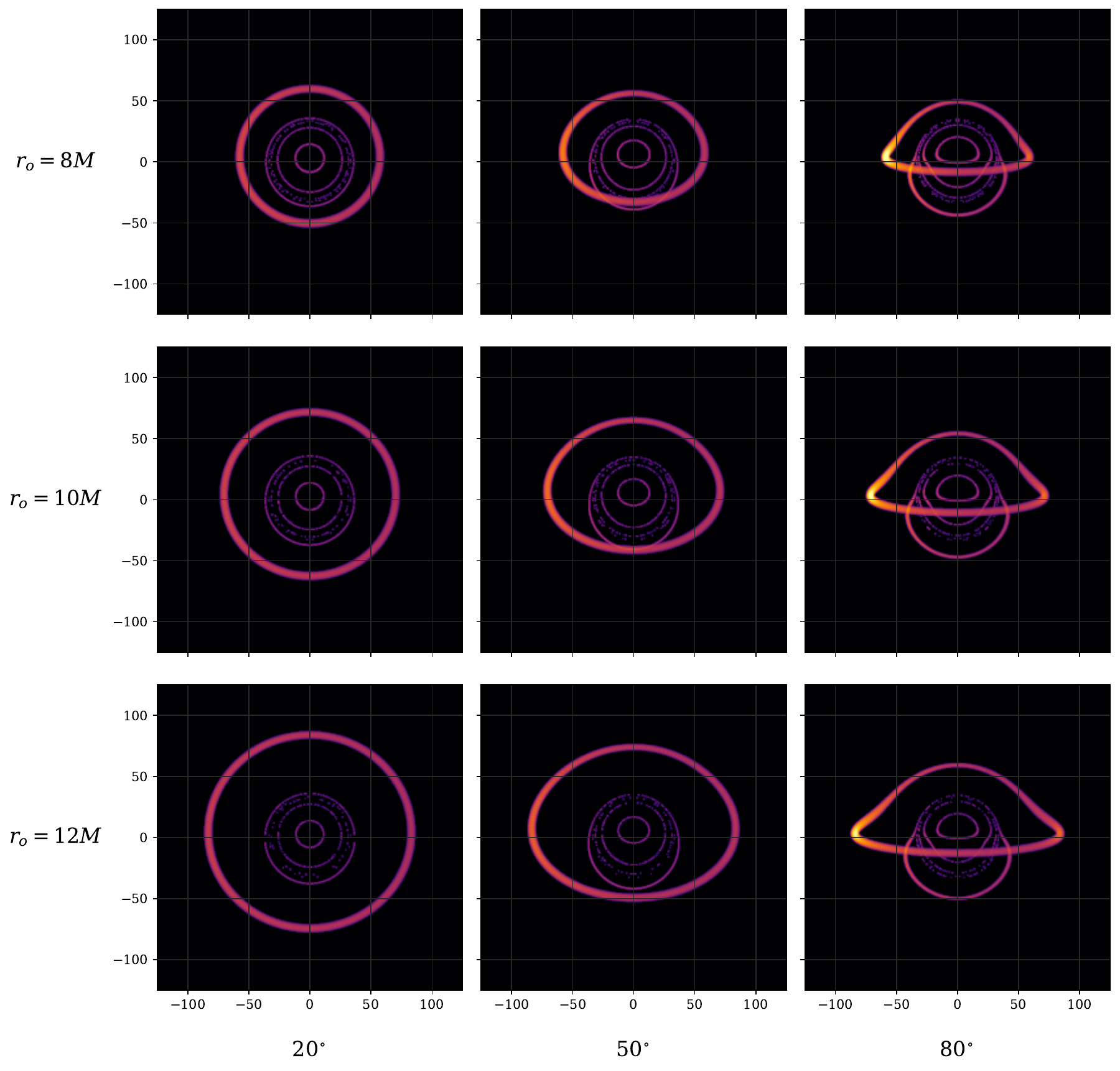}
    \caption{The integrated flux of the model M2 - $r_\Sigma = R = 2.5M$, for orbital radii $r_o\in\{8M,10M,12M\}$ (increasing from top to bottom) and inclination angles $i=20^{\circ},50^{\circ},80^{\circ}$ (increasing from left to right). The $x$- and $y$-axis are in units $\mu a s$. In addition to the Schwarzschild BH tracks, novel trajectories appear in the interior due to the stable LR in the interior. The model represents an ultra-compact horizonless object with a stable and unstable LR pair.}
    \label{fig:2p5}
\end{figure*}

\begin{figure*}
    \centering
    \includegraphics[width=\textwidth]{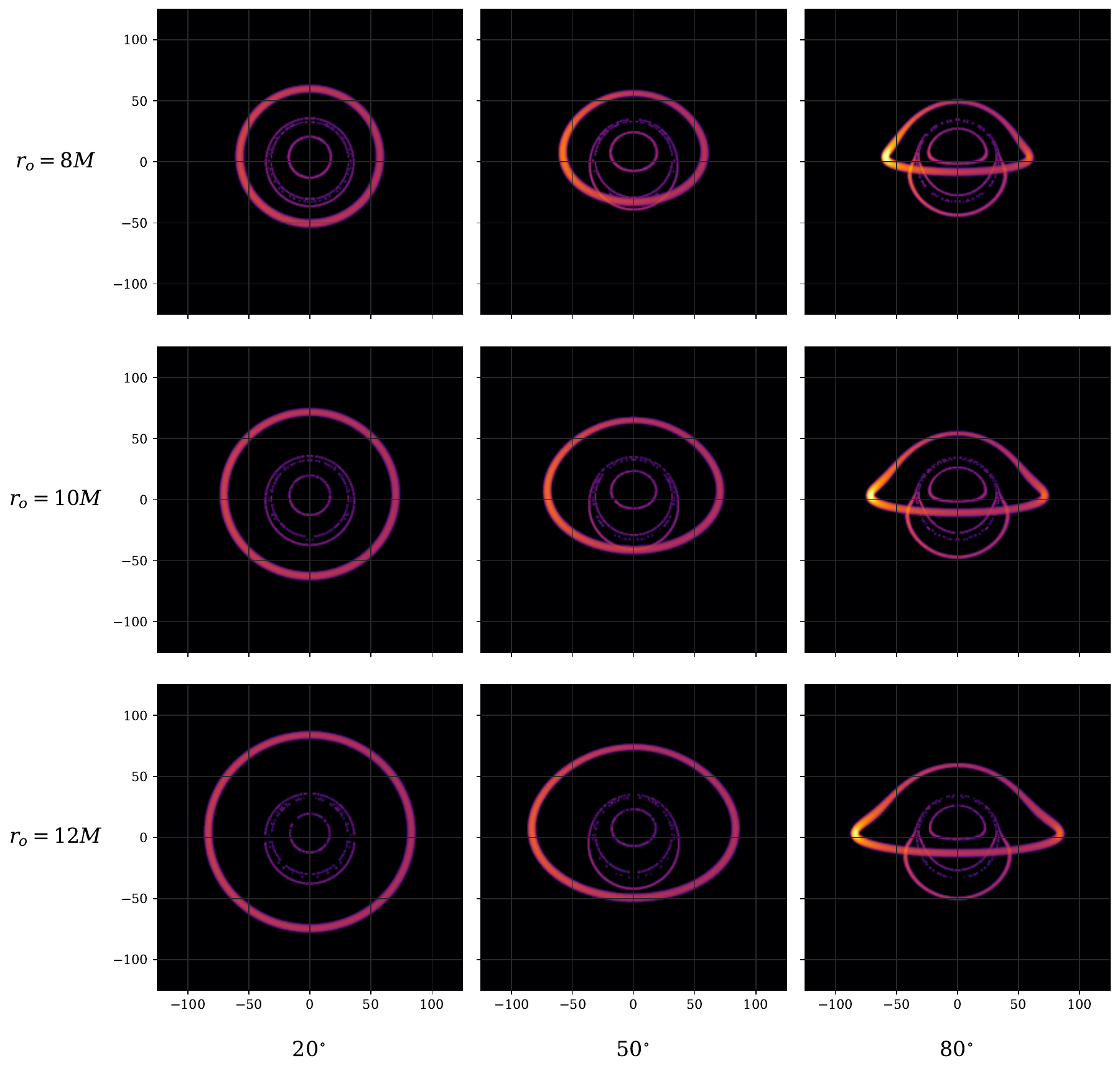}
    \caption{The integrated flux of the model M3 - $r_\Sigma = R = 3M$ for orbital radii $r_o\in\{8M,10M,12M\}$ (increasing from top to bottom) and inclination angles $i=20^{\circ},50^{\circ},80^{\circ}$ (increasing from left to right). The $x$- and $y$-axis are in units $\mu a s$. The image shows novel tracks in the interior, although the number of these new tracks is smaller than for the model M2. The model represents an ultra-compact horizonless object with a degenerate LR pair.}
    \label{fig:3}
\end{figure*}

\begin{figure*}
    \centering
    \includegraphics[width=\textwidth]{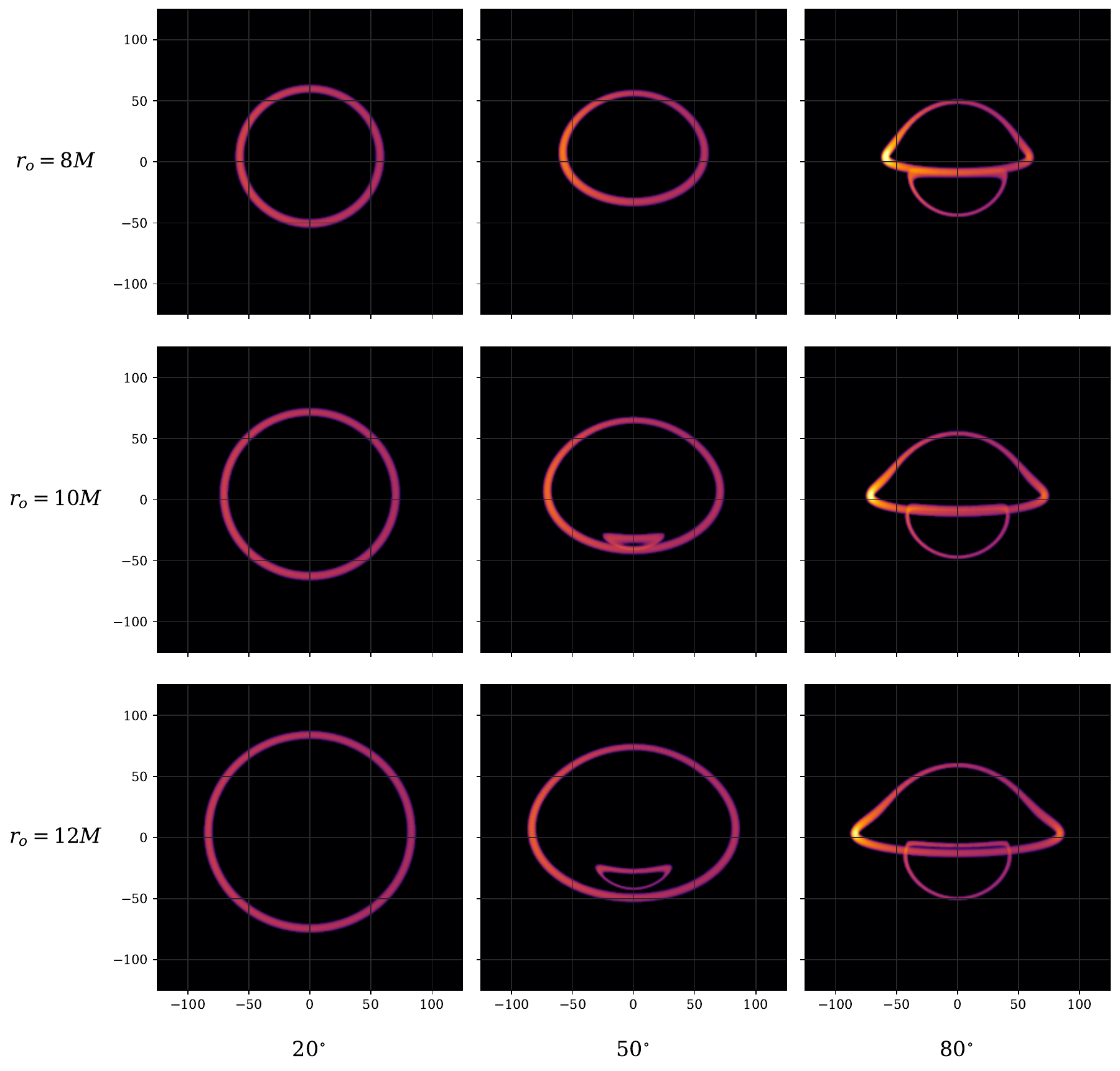}
    \caption{The integrated flux of the model  M4 - $r_\Sigma = R = 5M$, representing an ultra-compact  for orbital radii $r_o\in\{8M,10M,12M\}$ (increasing from top to bottom) and inclination angles $i=20^{\circ},50^{\circ},80^{\circ}$ (increasing from left to right). The $x$- and $y$-axis are in units $\mu a s$. The model resembles a BS with a melded secondary and plunge-through track.}
    \label{fig:5}
\end{figure*}

The Buchdahl case M1 is depicted in Fig.\ \ref{fig:2p25}, the ultra-compact models M2 and M3 are shown in Figs.\ \ref{fig:2p5} and \ref{fig:3}, respectively, and the non-ultra compact model M4 is depicted in Fig\ \ref{fig:5}. In these images, the impact of the orbital distance $r_o$ is well illustrated: increasing the distance greatly increases the size of the lensed image. As expected, changing the orbital distance mostly affects the primary track; there are also slight changes in the size of the secondary image, while the LR tracks stay virtually unchanged. Thus, if one were to consider a very low resolution observation, distinguishing primary tracks and secondary tracks might prove to be an easier task, if the hot spot were orbiting at greater distances.

\pagebreak


\end{document}